\documentstyle[prl,twocolumn,epsfig,aps]{revtex}

\newcommand{\be}{\begin{equation}} 
\newcommand{\en}{\end{equation}}
\newcommand{\bea}{\begin{eqnarray}}
\newcommand{\ena}{\end{eqnarray}}
\newcommand{\hbo}{\hbox to 1 true cm {\hfill } } 
\newcommand{\tr}{\hbox{tr}}

\begin{document} 

\title{ The Stefan-Boltzmann law: $SU(2)$ versus $SO(3)$ lattice gauge 
        theory } 

\author{ Kurt Langfeld, Hugo Reinhardt } 
\address{ Institut f\"ur Theoretische Physik, Universit\"at 
   T\"ubingen, D--72076 T\"ubingen, Germany }

\date{UNITU-THEP-17/99, \today}
\maketitle

\begin{abstract}
We investigate the high temperature 
limit of $SU(2)$ and $SO(3)$ lattice gauge theory, respectively. 
In particular, we study the Stefan-Boltzmann constant in both cases. 
As is well known, the Stefan-Boltzmann constant 
extracted from SU(2) lattice gauge theory by incorporating finite 
size effects is smaller than the continuum value which assumes three 
gluon degrees of freedom. On the other hand, the extrapolation of our 
$SO(3)$ lattice data comes much closer to the continuum value. This 
rises the question whether $SU(2)$ and $SO(3)$ lattice gauge theories 
represent different quantum theories in the continuum limit. 
\end{abstract}
\pacs{PACS: 11.10.Wx, 11.15.Ha, 12.38.Mh } 

Understanding the high temperature phase of Yang-Mills theory 
is essential for a wide span of physics, ranging from the evolution of 
the early universe and the description of compact 
stars~\cite{kol90}. With the advent 
of large scale numerical simulations of lattice gauge theories, 
it became evident that $SU(2)$ and SU(3) gauge theories undergo a 
phase transition at a critical temperature $T_c$ 
of a few hundred MeVs and that the 
high temperature phase is non-confining~\cite{kan98}. The fact that the 
effective (''running'') coupling constant becomes small at high energy 
scales in non-Abelian Yang-Mills theories indicates that the high 
temperature phase is described in terms of a gas of weakly interacting 
quarks and gluons forming a plasma. At temperatures well above the 
intrinsic energy scales, temperature is the only relevant 
scale. One therefore expects on general grounds that the vacuum 
energy density $\epsilon $ is related to the temperature $T$ by 
the Stefan-Boltzmann law 
\be 
\epsilon \; = \; \kappa \, T^4 \; , \hbox to 4cm {\hfill ($T$ large) } \; .
\label{eq:1} 
\en 
The Stefan-Boltzmann constant $\kappa $ only depends on the number 
of degrees of freedom constituting the high temperature phase. 
In the case of a pure SU(N) continuum gauge theory, a gas of 
$N^2-1$ gluons would imply $\kappa = (N^2-1) \, \pi^2 /15 . $ 

\vskip 0.3cm 
The remarkable finding of recent investigations of $SU(2)$ \cite{eng95} and 
SU(3)~\cite{eng90} pure gauge theory is that at temperatures $T= 2 \ldots 
3 T_c$ where the Stefan Boltzmann law is realized to good accuracy the ratio 
$\epsilon /T^4 $ strongly underestimates the asymptotic value $\kappa $ 
corresponding to a plasma made out of gluons. In particular for the 
$SU(2)$ case, one finds at twice the critical temperature that the ratio 
$\epsilon/T^4 $ only reaches 70\% of the asymptotic gluon plasma 
value~\cite{eng95}. 

\vskip 0.3cm 
A possible explanation of this discrepancy comes to mind: 
the continuum limit of $SU(2)$ lattice theory is not the same as the
usual continuum Yang-Mills theory defined in terms of the gauge connection.
In fact, lattice gauge theory is formulated in terms of link variables 
living in the gauge group, while the gauge potential of the continuum 
theory is defined in the algebra, which is the same for the SU(2) and 
the SO(3) group, respectively. Since furthermore the SU(2) and SO(3) 
lattice actions both reproduce the continuum action for zero lattice 
spacing, one would therefore expect that SU(2) and SO(3) lattice 
theory approach the same fix-point in the continuum limit. However, 
to our knowledge there is no rigorous proof that this is indeed the case. 
By contrast, since $SU(2) \simeq SO(3) \times Z_2 $, in addition to the 
$SO(3)$ degrees of freedom, the $SU(2)$ lattice theory contains $Z_2$ 
center degrees of freedom which, in principle, could survive the 
continuum limit and hence contribute to the Stefan-Boltzmann constant. 
The fact that the discrete degrees of freedom of a $Z_2$ theory can 
contribute to physical quantities is observed in the so-called Maximal 
Center Gauge~\cite{la98}: the effective $Z_2$ gauge theory is determined 
from the full SU(2) gauge theory by center projection~\cite{deb98} and 
is formulated in terms of vortices. It was observed that these 
vortices survive the continuum limit~\cite{la98} and are relevant 
infrared degrees of freedom. In fact, if these vortices are eliminated 
by hand, quark confinement~\cite{deb98,la98,la99a} and spontaneous breaking 
of chiral symmetry~\cite{for99} are lost. 

\vskip 0.3cm 
In this letter, we study the Stefan-Boltzmann constant in $SU(2)$ and 
$SO(3)$ lattice theories. The Stefan-Boltzmann constant measures 
the number of degrees of freedom forming the heat bath. Since the 
continuum extrapolation of $SO(3)$ lattice gauge theory can be formulated 
employing three gluon fields as in the case of continuum Yang-Mills theory, 
we expect that their Stefan-Boltzmann constant match, while a deviation 
from the continuum value should occur for the SU(2) case if 
center degrees of freedom survive at the continuum fixed point. 

\vskip 0.5cm 
The degrees of freedom of $SU(2)$ lattice gauge theory are 
defined by the link variables $U_\mu (x) = Z_\mu (x) O_\mu (x) $ 
while the link variables of $SO(3)$ lattice gauge theory, $O_\mu (x)$, 
can be constructed~\cite{bha81} from the link variables $U_\mu (x)$ by 
enforcing the constraint 
$Z_\mu (x)=1$, i.e., for the SO(3) case the link variables $U_\mu (x)$ 
are restricted to $ \tr U_\mu (x) > 0$. The actions for 
$SU(2)$ and $SO(3)$ gauge theories are given in terms of the 
plaquette variables 
\be 
S_{su2 / so3 } \; = \; \sum _{\mu > \nu, \{x\} } \beta _{F/A} 
P^{F/A} _{\mu \nu }(x) \; , \hbo 
\label{eq:2} 
\en 
where 
\bea 
P^F_{\mu \nu }(x) &:=& \frac{1}{2} \tr \biggl[ U_\mu (x) U_\nu (x+\mu ) 
U^\dagger _\mu (x+\nu ) U^\dagger _\nu (x) \, \biggr] \; . 
\label{eq:3} \\ 
P^A _{\mu \nu }(x) &=& \frac{4}{3} \biggl( P^F _{\mu \nu }(x) 
\biggr)^2 
\nonumber 
\ena 
The $SU(2)$ action is the standard Wilson action while the $SO(3)$ action 
is a special case of the Bhanot-Creutz action~\cite{bha81}. 
Either gauge theory is defined by its partition function 
\be 
Z_{su2 / so3 } \; = \; \int {\cal D} U_\mu (x) \; \exp \biggl\{ 
S_{su2 / so3 } \biggr\} \; . 
\label{eq:4} 
\en 
Finite temperature simulations can be performed by using asymmetric 
lattices with $N_\tau $ and $N_\sigma $ lattice points in time and 
spatial directions, respectively. The actual temperature is 
given by $T= 1/N_\tau a(\beta )$ where $a(\beta )$ is the lattice spacing. 
In order to detain the Casimir effect from distorting the energy 
density, a sufficiently large ratio $N_\sigma / N_\tau $ must 
be chosen. It was found in~\cite{eng95} that $N_\sigma / N_\tau = 4$ 
already yields reasonable results for $N_\tau \ge 4 $ and $\beta \le 2.8$. 

\begin{figure}[t]
\centerline{ 
\epsfxsize=8cm
\epsffile{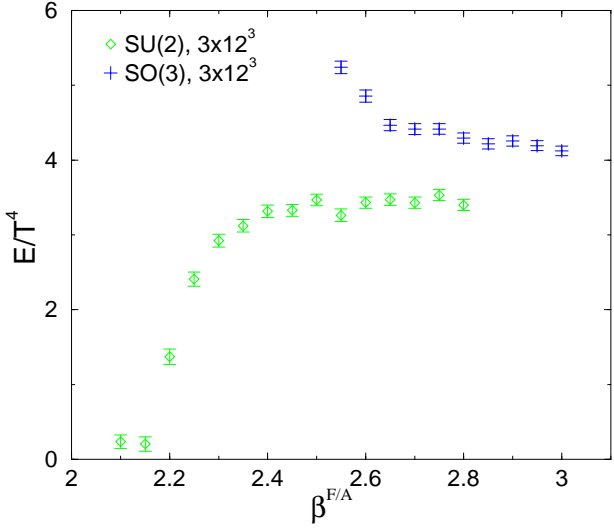}
}
\caption{ The raw data $E/T^4$ (\protect{\ref{eq:8}}) as function of 
   $\beta ^{F/A}$ for the $SU(2)$ and $SO(3)$ gauge theory, respectively, 
   for a $3\times 12^3$ lattice. } 
\label{fig:1} 
\end{figure}
\begin{figure}[t]
\centerline{ 
\epsfxsize=8cm
\epsffile{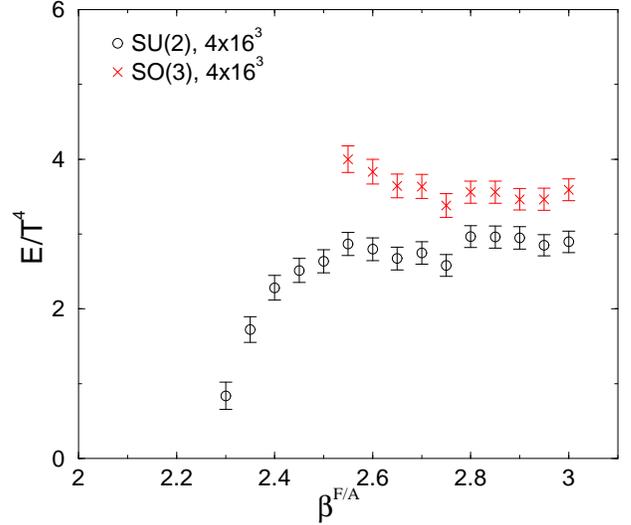}
}
\caption{ Same as figure 1 for a $4\times 16^3$ lattice. } 
\label{fig:1a} 
\end{figure}
\vskip 0.3cm 
An explicit expression for the energy density in terms of lattice 
variables can be found in the literature~\cite{eng95,eng90,hel85}. 
One finds 
\bea 
& & \frac{ \epsilon ^{su2/so3}}{ T^4 } = 3 \, \beta^{F/A} \, N_\tau ^4 \; 
\Biggl\{ 
\label{eq:5}  \\ 
& & 
\left[ 1 \, - \, \frac{ f^{F/A}_1(\beta^{F/A} ) }{ \beta^{F/A} } 
\right] \; \left( P^{F/A}_\tau \, - \, P^{F/A}_\sigma \right) 
\nonumber \\ 
&-& \beta^{F/A} \, f^{F/A}_2(\beta^{F/A} ) \, \left[ 2 P^{F/A}_0 \, - \, 
\left( P^{F/A}_\tau + P^{F/A}_\sigma \right) \right] \, \Biggr\} \; , 
\nonumber 
\ena 
where $P_\tau $ and $P_\sigma $ denote the expectation values of temporal 
and spatial plaquettes (in the asymmetric lattice) and 
$P^{F/A}_0$ is the plaquette expectation value for the symmetric lattice 
$N_\tau = N_\sigma $. The important observation is that the functions 
$f^{F/A}(\beta^{F/A} )$ approach finite values in the continuum limit 
$\beta \rightarrow \infty $. In particular, the function $f^F(\beta^F )$ 
can be deduced for several $\beta^F $ values from data reported 
in~\cite{eng95}. The function $f^A(\beta^A )$ can be calculated for 
$\beta ^A \gg 1$ by expanding the link variables $U_\mu (x) = \exp \{ i 
A_\mu (x) a \}$ in powers of the lattice spacing $a$ around the 
unit element. The result 
of this calculation can be also found in~\cite{eng95} and is referred to 
as the ''weak coupling regime'' of $SU(2)$ gauge theory. Note, however, 
that this calculation, which relies on the expansion of the link variables 
near $U_\mu =1$, is only justified for the $SO(3)$ case where 
the link variables are sufficiently close to the unit element for 
large $\beta ^{(A)}$. This is because in the $SU(2)$ case, 
this calculation does not properly take into account the non-trivial 
center elements $Z_\mu (x)$ which we consider the progenitor 
of vortices (see discussion below). In the continuum limit, 
one finally obtains $ \epsilon ^{su2/so3}  = E^{su2/so3}(\beta ^{F/A} 
\rightarrow \infty )$ where 
\be 
E^{su2/so3}/ T^4  \; = \; 
3 \, \beta^{F/A} \, N_\tau ^4 \; 
\left( P^{F/A}_\tau \, - \, P^{F/A}_\sigma \right) \; . 
\label{eq:8} 
\en 
In fact, one observes that the term in (\ref{eq:5}) proportional to 
$f_2(\beta )$ exponentially 
decreases with increasing $\beta ^F $ in the $SU(2)$ case and is for 
$\beta ^F>2.7$ orders of magnitude smaller than the dominant term 
proportional to $P_\tau - P_\sigma $. 
However, one observes significant corrections to (\ref{eq:8}) 
from the term proportional to $f_1^F( \beta _F )$ if $\beta ^F \in [2.5,3]$. 
Thus a suitable approximation to the Stefan-Boltzmann constant is 
\bea 
\kappa &:=& 
\frac{ \epsilon ^{su2/so3}}{ T^4 } 
\label{eq:9} \\ 
&\approx & 
3 \, \beta^{F/A} \, N_\tau ^4 \; 
\left[ 1 \, - \, \frac{ f^{F/A}_1(\beta^{F/A} ) }{ \beta^{F/A} } 
\right] \; \left( P^{F/A}_\tau \, - \, P^{F/A}_\sigma \right) \; . 
\nonumber 
\ena 

\begin{figure}[t]
\centerline{ 
\epsfxsize=8cm
\epsffile{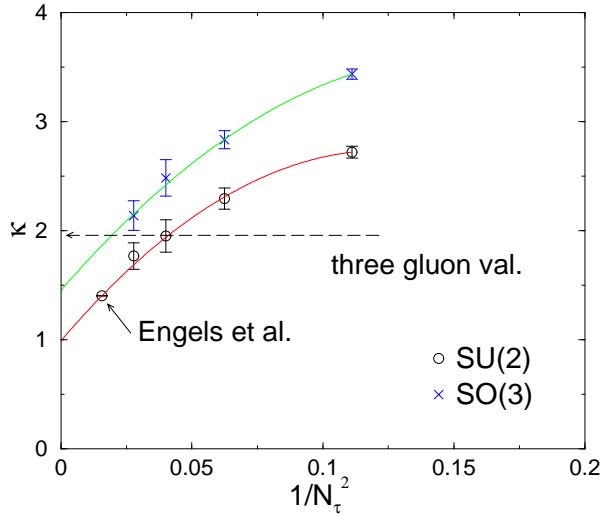}
}
\caption{ The Stefan-Boltzmann constant $\kappa $ (\protect{\ref{eq:9}}) 
   as function of $1/N^2_\tau $ for fixed ration $N_\sigma / N_\tau =4$. }
\label{fig:2} 
\end{figure}
\vskip 0.3cm 
Our numerical data for the $SU(2)$ case were obtained using the standard 
algorithm proposed by Creutz~\cite{creu80} while a novel heat bath algorithm 
was used for the study of $SO(3)$ gauge theory~\cite{la00}. Figures 
\ref{fig:1} and \ref{fig:1a} show the raw data $E/T^4$ as function of 
$\beta ^{F/A}$. One observes a clear signal of the deconfinement 
phase transition in the $SU(2)$ case. Since the $SO(3)$ lattice gauge theory 
possesses an un-physical phase transition at $\beta ^A =2.5 $ (independent 
of the lattice size)~\cite{bha81}, only data corresponding to the 
physical regime $\beta ^A >2.5$ are shown. In either case, a plateau 
value seems to be reached for $\beta ^{A/F} >2.7$. 

\vskip 0.3cm 
For an investigation of the Stefan-Boltzmann constant in either field 
theory (being defined as the continuum limit of the lattice formulation), 
a thorough study of the limit $N_\tau \rightarrow \infty $ is requested. 
Assuming eq.(\ref{eq:9}) and that 
the plateau value is reached for $\beta ^{F/A}=2.8$, we studied the 
$N_\tau $ dependence of $\kappa $ for fixed ratio $N_\sigma / N_\tau 
=4 $. We estimated $ f^F_1(\beta ^F) / \beta^F = 0.2 $ at 
$\beta ^F = 2.8$ with the help of the data tabulated in~\cite{eng95}. 
Since a detailed study of the non-perturbative $\beta $-function 
is not available for the $SO(3)$ case so far, we assume that 
the finite $\beta ^A$ correction to the continuum result is of the 
same order of magnitude (as suggested by lattice perturbation theory) 
and approximate $ f^A_1(2.8) \approx  f^F_1(2.8) $. 
Our numerical data 
are presented in figure \ref{fig:2}. We note that this approximation 
can lead to an relative error of 10\% for the absolute value of $\kappa $. 
Also shown 
is one data point for the $SU(2)$ case at $N_\tau = 8$ which is constructed 
with the help of the tabulated values in~\cite{eng95}. For 
guiding the eye, we have fitted the data points to the ansatz 
\be 
\kappa \; = \; \kappa _{\infty } \; + \; \frac{ c_1 }{ N_\tau ^2} 
\; + \; \frac{ c_2 }{ N_\tau ^4} \; , 
\label{eq:9a} 
\en 
which was investigated in~\cite{eng95}. 
Note that $\kappa _{SO(3)} / \kappa _{SU(2)}$ is 
insensitive to the absolute values of functions $f_1^A(\beta ^A)$ and 
$f_1^F(\beta ^F)$ as long as $f_1^A(\beta ^A) \approx f_1^F(\beta ^F)$. 
Our result for this ratio is tabulated in table~\ref{tab:1}). 

\vskip 0.3cm
Our results indicate that the Stefan-Boltzmann constant which emerges 
from the continuum extrapolation is larger in the $SO(3)$ than in the 
$SU (2)$ case. 
The explanation at hand is that in the $SU(2)$ case certain correlations 
survive even in the high temperature phase and prevent gluonic degrees of 
freedom from contributing to the Stefan-Boltzmann constant. 
In fact, lattice calculations performed in the Maximum Center 
Gauge~\cite{deb98} show that center vortices percolate in the confined 
phase implying strong gluonic correlations. As a result, the energy density 
vanishes in this regime. Furthermore, vortex dominance for the string tension 
is not only observed in the confinement regime, but also above the 
deconfinement phase transition for the spatial string tension~\cite{la99a}. 
In the deconfined phase, vortices partially align along the time axis 
but are still percolating in the 3-dimensional spatial universe 
resulting in a spatial string tension which is even larger than the 
string tension at zero temperature. This vortex scenario is compatible 
with dimensional reduction~\cite{app81} which support strong 
correlations in SU(2) 
lattice gauge theory even in the high temperature limit, thus 
effectively reducing the number of degrees of freedom participating 
in the gluonic heat bath. 

\begin{table} 
\caption{ The ratio of the estimates for the \\ 
   Stefan-Boltzmann constants of $SO(3)$ and $SU(2)$ gauge theory. } 
\label{tab:1}
\vskip 0.3cm
\begin{center} 
\begin{tabular}{ccccc}
$N_\tau \times N_\sigma ^3 $ & $3 \times 12^3 $ & $4 \times 16^3$ & 
$5\times 20^3 $ & $6 \times 24^3$ \\[0.5ex] \hline
$\kappa _{SO(3)} / \kappa _{SU(2)} $ & $1.26 \pm 0.03 $ & 
$1.24 \pm 0.07 $ & $1.27 \pm 0.08 $ & $1.21 \pm 0.1 $ 
\end{tabular} 
\end{center}
\end{table} 
\vskip 0.3cm
In conclusions, we have studied for the first time the Stefan-Boltzmann 
constant of $SO(3)$ gauge theory by an extrapolation of lattice Monte-Carlo 
data to the continuum and infinite volume limit. We find preliminary 
evidence that this constant is about 20\% larger than the 
corresponding constant of $SU(2)$ gauge theory. Given the fact that the 
Stefan-Boltzmann constant which arises from the continuum 
extrapolation of SU(2) lattice gauge theory is roughly 30\% smaller than 
the expectation provided by three gluon degree's of 
freedom~\cite{eng95}, our results indicate that the Stefan-Boltzmann 
constant of $SO(3)$ gauge theory comes closer to the continuum 
expectation than the $SU(2)$ one. 
In our opinion, a large scale numerical analysis 
(comparable with~\cite{eng95}) of the $SO(3)$ theory is highly desirable 
for a more detailed study of the important question whether $SO(3)$ and 
$SU(2)$ lattice gauge theories give rise to {\it different} continuum 
field theories. 

\vskip 0.3cm
{\bf Acknowledgements:} Helpful discussions with M.~Ilgenfritz are greatly 
acknowledged. We thank M.~Engelhardt and R.~Alkofer for comments 
on the manuscript. This work is supported in part by Deutsche 
Forschungsgemeinschaft under contract DFG-Re 856/4-1.

\end{document}